\pdfoutput=1
\documentclass[conference]{IEEEtran}
\IEEEoverridecommandlockouts % Needed for the license information in the footnote
\usepackage{cite}
\usepackage{amsmath,amssymb,amsfonts}
\usepackage{graphicx}
\usepackage{textcomp}
\usepackage{xcolor}
\usepackage{algorithm}
\usepackage{algpseudocode}
\def\BibTeX{{\rm B\kern-.05em{\sc i\kern-.025em b}\kern-.08em
		T\kern-.1667em\lower.7ex\hbox{E}\kern-.125emX}}
\usepackage{hyperref}

\begin{document}

\title{Deep Ocean\\
	\large A blockchain-agnostic dark pool protocol
	\thanks{This work is licensed under the Creative Commons Attribution-ShareAlike 4.0 International License.}
}

\author{
	\IEEEauthorblockN{Bruno Fran\c{c}a}
	\IEEEauthorblockA{
		Trinkler Software\\
		\textit{company@trinkler.software}}
	\\\\Version 1 -- February 19, 2019
}

\maketitle
\thispagestyle{plain} %to have page numbers
\pagestyle{plain} %to have page numbers

\begin{abstract}
We introduce a new cryptographic protocol, called Deep Ocean, that implements a blockchain-agnostic dark pool for cryptocurrencies. Deep Ocean is a layer-two protocol, meaning that it can work with any two cryptocurrencies, as long as there exists an underlying settlement mechanism, for example performing atomic swaps.
\end{abstract}

\section{Introduction}
In finance, a dark pool is a private exchange for trading financial instruments such as securities. It differs from other exchanges in that its order is not public, only the exchange operators have access to it. Thus it is not possible to see how much liquidity exists on either side of the market.

The name comes from the fact that price discovery is not possible in dark pools. Since the trades are not revealed until some time after their execution, the market is no longer transparent.

Dark pools are especially attractive to traders who want to buy or sell large blocks of securities. Performing these large trades in an open exchange will cause market impact and invite front-running from other exchange participants, thus increasing the cost of the trade. When trading in dark pools, traders do not need to worry about these problems since their order will not be known to anyone except the exchange operators.

Another frequent user of dark pools are hedge funds, who wish to be able to trade without \textit{'showing their hand'} to other traders. Dark pools either do not release their trade history or do so with as much delay as legally allowed.

All these problems also exist in cryptocurrency exchanges, thus a need exists for dark pools in the cryptocurrency world. Deep Ocean is such a dark pool. It is a protocol that is executed between an exchange operator and its users, allowing the exchange to maintain an order book without actually learning the order sizes. Furthermore, it is agnostic to the settlement method so, when used with atomic swaps, it allows for non-custodial trades.

\section{Overview}\label{sec:overview}
Deep Ocean can work with any pair of blockchains, as long as there is some way of settling the trade. It can be by an atomic swap, or by a decentralized exchange or even by using a centralized exchange. The only function of Deep Ocean is to allow two parties to agree on the details of their trade without revealing \textit{a priori} their order sizes.

In order to do this, Deep Ocean requires a third-party. The third-party is only needed to relay messages between the two parties and to punish them if they don't follow the protocol. It is impossible for the third-party to learn the order size, or any other private information, of any one of the parties so, the third-party does not need to be trusted.

The role of the third-party will normally be played by an exchange operator, and the other two parties will be exchange users. In this setup, if one of the users deviates from the protocol, the exchange operator will expel him from the dark pool. This forces the users to follow through with the trade after they start the protocol.

Deep Ocean follows the design of older dark pools and matches all orders at market price. This means that the exchange operator also has the responsibility of providing the current market price to the dark pool users \footnote{Since the users can easily verify the current price by themselves, for example by checking that the price is consistent with the price reported by coinmarketcap.com, they do not need to trust the exchange operator.}. This method results in a faster and more liquid exchange, since all orders can be matched against each other, which is crucial because most users of dark pools prefer liquidity and speed over price \footnote{It is also possible to modify the Deep Ocean protocol in order to allow limit orders, and consequently matching at prices different than market price. This modification will also hide the limit prices of each order, in addition to their size. However, this will result in a slower dark pool.}.

Let us imagine three parties: Alice and Bob, who are users of the dark pool, and Eve, who is the dark pool operator. All messages between Alice and Bob go through Eve. We assume that Alice and Bob may be malicious and Eve is honest-but-curious, meaning that she follows the protocol but will try to get any information that she can. The Deep Ocean protocol then works as follows:

\begin{enumerate}
	\item Both Alice and Bob register with Eve to become valid users. This is an one-time step and can be something as simple as providing an email address or more involved like a KYC/AML registration process.
	\item Both Alice and Bob send their orders to Eve. The orders consist of the asset they wish to buy, the asset which they wish to sell and, optionally, the limit price.
	\item Eve then tries to match buyers and sellers. If the current market price exceeds the limit price, the order isn't considered. When a match is found, Eve sends a message to both parties stating that they were matched at the current market price. At this point, both parties need to confirm that they wish to proceed with the trade. If they confirm, they must complete the trade or be punished.
	\item Alice and Bob then run a secure comparison protocol \cite{brandt2005efficient} between themselves to find who has the smallest order size without revealing their respective order sizes. Since every message between the two is signed and has accompanying zero knowledge proofs, it is easy to prove that malicious behavior has occurred. Eve then punishes the offending party, maybe by expelling him from the dark pool or by charging a fine.
	\item After running the secure comparison protocol, Alice and Bob will know who has the smallest order. Let us imagine it is Bob. Eve now can delete Bob's order from her book and she can begin to try to find another match for Alice, since part of Alice's order will not be filled in this trade.
	\item Bob now reveals his order size to Eve and Alice. There is no problem with this since he expects to have his order completely filled in the next few steps, otherwise Alice will be punished. No one knows how much of Alice's order remains unfilled.
	\item Now Alice, Bob and Eve can start the trade settlement. They already know the price and the size, so they only need to exchange the rest of the information (addresses, fees, etc).
\end{enumerate}

Deep Ocean has several advantages over regular cryptocurrency exchanges:
\begin{itemize}
	\item It is blockchain agnostic so it can be used with any blockchain that supports atomic swaps.
	\item The exchange operator only needs to be trusted to maintain the list of dark pool users and relay the messages.
	\item Users only reveal their order sizes when they know that their order is going to be completely filled.
\end{itemize}

The only disadvantage is that it requires users to be online during the whole process. Although the comparison protocol is relatively fast, users may have to wait some time until they find a counterparty for their trade.

\section{Notation}
$\mathbb{Z}_p$ is the integer field of order p. So the integers from \textit{0} (inclusive) to \textit{p} (exclusive). All operations on fields will be of modular arithmetic. $\mathbb{Z}_p^*$ is the integer field of order \textit{p} except \textit{0}. Lower case letters will represent integers, upper case letters will represent elliptic curve points. $H()$ represents a hash function, it takes some data (integers, EC points, etc) and transforms it into an integer.

$\leftarrow$ means assignment. Ex: $x \leftarrow 0$ means "assign the value 0 to x". $\xleftarrow{R}$ means "pick at random from". Ex: $z \xleftarrow{R} \mathbb{Z}_p$ means "pick a random number from $\mathbb{Z}_p$ and assign to z".

We denote the generator element of an elliptic curve as $G$ and its neutral element as $O$. The neutral element is the single point that obeys the following two properties:

\begin{itemize}
	\item $\forall P \in \mathbb{G}_p: P+O=P$
	\item $\forall P \in \mathbb{G}_p: P-P=O$
\end{itemize}

\section{Cryptographic building blocks}

\subsection{EC Pedersen commitments}
A commitment scheme allows one to commit to a chosen value while keeping it hidden to others, with the ability to reveal the committed value later. To do an EC Pedersen commitment you just need to know two generators $G_1, G_2$ of an elliptic curve.

\begin{itemize}
\item \textbf{Commitment:} Given a value $x \in \mathbb{Z}_p$, choose $r \xleftarrow{R} \mathbb{Z}^*_p$ and calculate $C=xG_1+rG_2$. The commitment is $C$.
\item \textbf{Reveal:} To reveal the commitment just send $x$ and $r$. The verifier then checks that $C=xG_1+rG_2$.
\end{itemize}

\subsection{EC ElGamal}
This a variant of ElGamal encryption. For this scheme we only need an elliptic curve and the corresponding generator $G$.

\begin{itemize}
\item \textbf{Key generation:} Choose $x \xleftarrow{R} \mathbb{Z}_p$ and calculate $P=xG$. $x$ is the private key and $P$ is the public key.
\item \textbf{Encryption:} Given a message $m \in \mathbb{Z}_p$, choose $r \xleftarrow{R} \mathbb{Z}^*_p$ and calculate $A=mG+rP$ and $B=rG$. The cyphertext is $(A,B)$.
\item \textbf{Decryption:} Calculate $A-xB=(mG+rP)-x(rG)=(mG+rP)-rP=mG$.
\end{itemize}

Note that after decryption we only know $mG$ and not $m$. To get $m$ we need to break the elliptic curve discrete logarithm problem. Because of this, this variant of ElGamal can only be used when $m$ is a small number (less than 32 bits). Thankfully, all messages that we'll need to decrypt during this protocol are only 1-bit long.

\subsection{Proof of knowledge of logarithm}
This is a zero knowledge proof that you know an integer $x$ such that $P=xG$, where $G$ is a generator of an elliptic curve.

\textbf{Public input:} $P,G$

\textbf{Private input:} $x$

\textbf{Prover:}
\begin{enumerate}
\item Choose $z \xleftarrow{R} \mathbb{Z}^*_p$ and calculate $A=zG$.
\item Calculate $c=H(A), \ c \in \mathbb{Z}_q$.
\item Calculate $r=z+cx \mod q$
\end{enumerate}

\textbf{Proof:} $A, r$

\textbf{Verifier:}
\begin{enumerate}
\item Check that $rG=A+cP$
\end{enumerate}

\subsection{Proof of equality of logarithms}
This is a zero knowledge proof that you know an integer $x$ such that $P=xG_1$ and $Q=xG_2$, where $G_1$ and $G_2$ are generators of an elliptic curve.

\textbf{Public input:} $P,G_1,Q,G_2$

\textbf{Private input:} $x$

\textbf{Prover:}
\begin{enumerate}
\item Choose $z \xleftarrow{R} \mathbb{Z}^*_p$. Calculate $A=zG_1$ and $B=zG_2$.
\item Calculate $c=H(A,B), \ c \in \mathbb{Z}_q$.
\item Calculate $r=z+cx \mod q$
\end{enumerate}

\textbf{Proof:} $A, B, r$

\textbf{Verifier:}
\begin{enumerate}
\item Check that $rG_1=A+cP$ and $rG_2=B+cQ$
\end{enumerate}

It is trivial to extend this algorithm to prove logarithm equality between more than two elliptic curve points.

\subsection{Proof that an encrypted value is either 0 or 1}
This is a zero knowledge proof that an EC ElGamal cyphertext $(C,D)=(mG+rP,rG)$ encrypts either 0 or 1 so, $m \in \{0,1\}$.

\textbf{Public input:} $(C,D), P, G$

\textbf{Private input:} $m,r$

\textbf{Prover:}
\begin{enumerate}
\item \begin{itemize}
\item if m=0: Choose $r_1,d_1,w \xleftarrow{R} \mathbb{Z}^*_p$. Calculate $A_1=r_1G+d_1D$, $B_1=r_1P+d_1(C-G)$, $A_2=wG$ and $B_2=wP$.
\item if m=1: Choose $r_2,d_2,w \xleftarrow{R} \mathbb{Z}^*_p$. Calculate $A_1=wG$, $B_1=wP$, $A_2=r_2G+d_2D$ and $B_2=r_2P+d_2C$.
\end{itemize}
\item Calculate $c=H(A_1,B_1,A_2,B_2), \ c \in \mathbb{Z}_q$.
\item \begin{itemize}
\item if m=0: Calculate $d_2=c-d_1 \mod q$ and $r_2=w-rd_2 \mod q$.
\item if m=1: Calculate $d_1=c-d_2 \mod q$ and $r_1=w-rd_1 \mod q$.
\end{itemize}
\end{enumerate}

\textbf{Proof:} $A_1, B_1, r_1, d_1, A_2, B_2, r_2, d_2$

\textbf{Verifier:}
\begin{enumerate}
\item Check that $c=d_1+d_2 \mod q$, $A_1=r_1G+d_1D$, $B_1=r_1P+d_1(C-G)$, $A_2=r_2G+d_2D$ and $B_2=r_2P+d_2C$.
\end{enumerate}

\subsection{Proof of shuffle}
A zero knowledge proof of shuffling is a proof that two vectors of cyphertexts, $\vec{A}=(A_1, A_2, \ldots, A_n)$ and $\vec{B}=(B_1, B_2, \ldots, B_n)$, encrypt the same plaintexts and that a permutation was applied, without revealing the actual permutation.

We will use a proof of shuffling proposed by Bayer and Groth \cite{bayer2012efficient}.

\section{Technical description}
Now we will describe the Deep Ocean algorithm of Section \ref{sec:overview} in more detail:

\begin{enumerate}
\item There is a server, called Eve, that maintains a list of all the current participants in the dark pool. It records their public keys. Alice $(i=1)$ and Bob $(i=2)$ are both part of the dark pool. All communication between Alice and Bob will happen through Eve.
\item Alice and Bob send their orders to Eve. The order is the tuple (asset to buy, asset to sell, limit price). In both cases the limit price is optional.
\item Let us assume that Eve matches Alice and Bob. To do this, she sends a message to both of them, containing each others public keys and the current market price. Alice and Bob check the  that the market price is correct.
\item Alice and Bob now confirm that they want to proceed with the trade. This is the last time that they are able to abort. If they abort later, they will be punished by Eve.
\item Alice and Bob create a EC keypair $(x_i, P_i)$. They send their public key to each other, together with a ZK proof of knowledge of logarithm. Finally, each one computes $P=P_1+P_2$.
\item Let us denote Alice's and Bob's order size as $s_i$. Let us denote the \textit{j}-th bit of $s_i$ as $b_{ij}$. Alice and Bob, for each bit $b_{ij}$, choose $r_{ij} \xleftarrow{R} \mathbb{Z}^*_p$.
\item Alice and Bob encrypt each bit $b_{ij}$ using EC ElGamal, $(A_{ij}, B_{ij})=(b_{ij}G+r_{ij}P, r_{ij}G)$. Alice and Bob, for all bits $b_{ij}$, send each other the corresponding encryption together with a ZK proof that it encrypts either a zero or a one.
\item Alice and Bob perform the following calculations, for all $j \in [1, \ldots , k]$:
\[V_j=G+A_{2j}-A_{1j}+\sum_{d=j+1}^{k} (2^d-2)(A_{1d}-A_{2d}) \]
\[U_j=B_{2j}-B_{1j}+\sum_{d=j+1}^{k} (2^d-2)(B_{1d}-B_{2d}) \]
\item Alice shuffles the $k$ vectors $(V_j, U_j)$ by the index $j$ and sends the shuffled vectors to Bob together with a ZK proof of correct shuffle.
\item Bob shuffles the $k$ vectors $(V_j, U_j)$ by the index $j$ and sends the shuffled vectors to Alice together with a ZK proof of correct shuffle.
\item Alice and Bob choose, for all $j$, $m_{ij} \xleftarrow{R} \mathbb{Z}^*_p$. Alice and Bob calculate, for all $j$, $V_j^i=m_{ij}V_j$ and $U_j^i=m_{ij}U_j$, and send it to each other together with a proof of logarithm equality.
\item Alice and Bob calculate, for all $j$, $W_j^i=x_i(U_j^1+U_j^2)$ and send it to each other together with a proof of logarithm equality.
\item Alice and Bob compute, for all $j$:
\[T_j = (V_j^1+V_j^2)-(W_j^1+W_j^2)\]
\item Alice and Bob, for all $j$, decrypt together all $T_j$. If, for any $j$, $T_j=O$, then $s_1>s_2$, meaning Alice's order is bigger than Bob's. Otherwise, $s_1 \leq s_2$, meaning Bob's order is bigger than Alice's. Let us assume that Bob's order is bigger.
\item Alice calculates $r_{i} = \sum_j=1^k 2^{j-1} r_{ij}$ and sends $s_i$ and $r_i$ to Bob.
\item Bob checks that $s_iG+r_iP=\sum_j=1^k 2^{j-1} A_{ij}$.
\item At this point Alice and Bob can begin settlement of the trade. The specifics depend on the method chosen.
\end{enumerate}

\section{Conclusion}
In this paper we showed how a dark pool can be implemented through a cryptographic protocol. Such a protocol allows non-custodial cryptocurrency trades that do not leak any information about order sizes without requiring a trusted server.

We hope that this protocol will bring more institutional investors and other large traders into the cryptocurrency market.

\bibliographystyle{IEEEtran}
\bibliography{../bibliography.bib}

\end{document}